\newlength{\absize}
\documentstyle[12pt]{article}
\renewcommand{\baselinestretch}{1.5}

\begin{document}
\thispagestyle{empty}
\pagestyle{empty}
\renewcommand{\thefootnote}{\fnsymbol{footnote}}
\newcommand{\starttext}{\newpage\normalsize
\pagestyle{plain}
\setlength{\baselineskip}{4ex}\par
\setcounter{footnote}{0}
\renewcommand{\thefootnote}{\arabic{footnote}}
}
%\fi

\newcommand{\figsize}{}

\newdimen\tdim
\tdim=\unitlength

\newcommand{\preprint}[1]{\begin{flushright}
\setlength{\baselineskip}{3ex}#1\end{flushright}}
\renewcommand{\title}[1]{\begin{center}\LARGE
#1\end{center}\par}
\renewcommand{\author}[1]{\vspace{2ex}{\Large\begin{center}
\setlength{\baselineskip}{3ex}#1\par\end{center}}}
\renewcommand{\thanks}[1]{\footnote{#1}}
\renewcommand{\abstract}[1]{\vspace{2ex}\normalsize\begin{center}
\centerline{\bf Abstract}\par\vspace{2ex}\parbox{\absize}{#1
\setlength{\baselineskip}{2.5ex}\par}
\end{center}}

%%%%%%%%%%%%%%%%%%%%%%%%%%%%%%%%%%%%%%%%%%%%%%%%%%%%%%%%%%%%%%%%%%%
%%%                   MACROS from lmacros                       %%%
%%%%%%%%%%%%%%%%%%%%%%%%%%%%%%%%%%%%%%%%%%%%%%%%%%%%%%%%%%%%%%%%%%%
%Move the definition below past the style to get (c.s.e) equations
%\def\theequation{\thesection.\arabic{equation}}
\newcommand{\segment}[2]{\put#1{\circle*{2}}}
\newcommand{\fig}[1]{figure~\ref{#1}}
\newcommand{\hc}{{\rm h.c.}}
\newcommand{\ds}{\displaystyle}
\newcommand{\eqr}[1]{(\ref{#1})}
\newcommand{\tr}{\mathop{\rm tr}}
\def\spur#1{\mathord{\not\mathrel{#1}}}
\def\lta{\mathrel{\displaystyle\mathop{\kern 0pt <}_{\raise .3ex
\hbox{$\sim$}}}}
\def\gta{\mathrel{\displaystyle\mathop{\kern 0pt >}_{\raise .3ex
\hbox{$\sim$}}}}
\newcommand{\sechead}[1]{\medskip{\bf #1}\par\bigskip}
\newcommand{\cross }{\hbox{$\times$}}
\newcommand{\ol}{\overline}
\newcommand{\bra}{\Bigl\langle}
\newcommand{\ak}{\Bigm|}
\newcommand{\ket}{\Bigr\rangle}
\newcommand{\Bra}[1]{\left\langle #1 \right|}
\newcommand{\Ket}[1]{\left| #1 \right\rangle}
\newcommand{\g}[1]{\gamma_{#1}}
\newcommand{\half}{{1\over 2}}
\newcommand{\cl}[1]{\begin{center} #1\end{center}}
\newcommand\etal{{\it et al.}}
\newcommand{\prl}[3]{Phys. Rev. Letters {\bf #1} (#2) #3}
\newcommand{\prd}[3]{Phys. Rev. {\bf D#1} (#2) #3}
\newcommand{\npb}[3]{Nucl. Phys. {\bf B#1} (#2) #3}
\newcommand{\plb}[3]{Phys. Lett. {\bf #1B} (#2) #3}
\newcommand{\ie}{{\it i.e.}}
\newcommand{\etc}{{\it etc.\/}}
\newcommand{\boxit}[1]{\ba{|c|}\hline #1 \\ \hline\ea}
\newcommand{\mini}[1]{\begin{minipage}[t]{20em}{#1}\vspace{.5em}
\end{minipage}}
%%%%%%%%%%%%%%%%%%%%%%%%%%%%%%%%%%%%%%%%%%%%%%%%%%%%%%%%%%%%%%%%%%%
%%%                 END MACROS from lmacros                     %%%
%%%%%%%%%%%%%%%%%%%%%%%%%%%%%%%%%%%%%%%%%%%%%%%%%%%%%%%%%%%%%%%%%%%

\newcommand{\Tr}{\mathop{\rm Tr}}

\preprint{\#HUTP-96/A041\\ 11/96}
\title{
A Little Large $N$ Group Theory
\thanks{Research
supported in part by the National Science Foundation under Grant
\#PHY-9218167.}
}
\author{Hael Collins and Howard Georgi \\
Lyman Laboratory of Physics \\
Harvard University \\
Cambridge, MA 02138 \\
}
\date{}
\abstract{
We discuss the group theory relevant to the ground-state baryons in large
$N_c$ QCD. For very large representation, the group generators become
classical variables. We find the form of the classical generators for the
completely symmetric $N$ index representation of $SU(m)$ as
$N\rightarrow\infty$ and derive an integral formula for the matrix elements
of an arbitrary polynomial in the group generators between low-spin baryon
states in the large $N$ limit.
}

\starttext
\setcounter{page}{1}

The idea of replacing the $SU(3)$ gauge symmetry of QCD with an
$SU(N)$ symmetry and studying the $N\to\infty$ limit, as articulated
by 't Hooft in \cite{thooft}, has led to an important qualitative
understanding of some of the properties of QCD, such as Zweig's rule
and the narrowness of resonances.  Witten \cite{witten} later provided
a conceptual framework to include baryons in the theory.  In
Witten's picture, a baryon is described by a Hartree-Fock equation with 
each quark moving in the mean potential generated by all the other 
quarks in the baryon.  Using this description for the baryons,
\cite{djm,georgi,luty} (see also \cite{classical}) one can show that to leading
order in $N$, the low-spin baryons have a
spin-flavor symmetry which we will denote as $SU(m)$, {\it
e.g.\/}\ for an $N_f$-flavor theory $m=2N_f$.  In this letter, we
study the group theory associated with the large representations of
this $SU(m)$.  In doing so, we develop an elegant integral formula for
matrix elements of $SU(m)$ generators between low-spin baryon states.
This formula provides an insight into the nature of the large $N$
enhancement of these matrix elements.

An $N$ quark baryon lives in the completely antisymmetric
representation of the color $SU(N)$.  An $s$-wave ground state, to
satisfy fermi statistics, must then be completely symmetric under the
$SU(m)$ symmetry.  Therefore, the representation of $SU(m)$ relevant
to large $N$ baryons is the completely symmetric combination of $N$
defining representations.  We will examine the matrix elements of the
group generators in this representation, which we will denote by
$T^\alpha_\beta$, to leading order in $N$, and
will discover that many properties of the generators can be obtained
rather simply in this limit.

Our primary tool will be the fact that for large $N$, the commutator
of two group generators is
lower order in $N$ than the product. The product is order $N^2$, while
the commutator is order $N$ (times the structure constants, which
depend on $m$ but not on $N$).  Thus to leading order in $N$, the
generators can be simultaneously diagonalized---they become
essentially classical variables~\cite{classical}~\cite{djm}~\cite{formal}.

Explicitly calculating the traces in the representation space of 
products of generators and taking the $N\to\infty$ limit leads to 
the following form for the generators:\footnote{We will 
present a detailed proof for this result in a future article.  The 
same result also emerges in the geometric quantization of $SU(m)$.
\cite{kirillov}}
\begin{equation}
\begin{array}{c}
T^\alpha_\beta\, | u \rangle = {\cal T}^\alpha_\beta(u)\, |u\rangle \\
{\cal T}^\alpha_\beta(u) = N \left( u^\alpha \bar u_\beta - {1\over m}
\delta^\alpha_\beta \right)
\end{array}
\label{u}
\end{equation}
where $u$ is an $m$ component complex unit vector and $\bar u_\beta
\equiv u^{\beta*}$.  Notice that this form for the generators 
has a single non-degenerate eigenvalue ($(m-1)/m$) and $(m-1)$ degenerate
eigenalues ($-1$ for the eigenvectors
orthogonal to $u$) and that moreover the overall
phase of $u$ does not matter.  The
generators in the large $N$ limit belong
to the classical coset space,
\begin{equation}
SU(m)/[SU(m-1)\times U(1)] \cong {\bf CP}^{m-1}\,,
\label{coset}
\end{equation}
but (\ref{u}) is stronger than (\ref{coset}), because the normalization is also
fixed.
For an $SU(2)$ symmetry, this limit reproduces the familiar statement
that angular momentum becomes a classical vector when the spin grows very 
large, since then $SU(2)/U(1)$ is a two-sphere that labels the direction of
the classical angular momentum.

We show in the following section how the form for the generators,
(\ref{u}), naturally arises in the $m$ dimensional harmonic
oscillator.  The harmonic oscillator energy eigenstate live in
precisely the same representation of $SU(m)$ as do the ground state
baryons in large $N$ QCD.  In section \ref{eigenstates}, we find the
eigenstates of the group generators, and in section \ref{spin}, we use
these eigenstates to find an integral formula for the leading
contribution to matrix elements of an arbitrary polynomial in the
generators between low-spin states.

\section{$m$ Dimensional Harmonic Oscillator\label{harmonic}}

We begin by analyzing the large $N$ limit of the $m$-dimensional
harmonic oscillator.  In addition to providing a familiar system which
has a well-known classical limit, this example will prove useful later
when we attempt to construct the low-spin baryons.  In the large $N$
limit of QCD, the important behavior of the quarks comprising a baryon
can be represented by {\it colorless, bosonic\/} creation and
annihilation operators, as has been applied in the ``quark
representation'' of \cite{djm} and in the analysis of \cite{luty}.
We will use the energy eigenstates of the harmonic oscillator as an 
intermediate step bridging between the baryon states and the $SU(m)$ 
generator eigenstates.

The $m$-dimensional harmonic oscillator corresponds to the Hamiltonian
(setting the spring constant and the mass to 1)
\begin{equation}
H=\half\left({\vec x}^{\,2}+{\vec p}^{\,2}\right)
=\sum_\alpha  {a^\alpha}^{\,\dagger} \, a_\alpha +{m\over2}
\end{equation}
where ${a^\alpha}^{\dagger} $ and $a_\beta $ represent the raising 
and lowering operators
\begin{equation}
a_\alpha ={x^\alpha -i\,p^\alpha \over\sqrt2}\quad\quad 
{a^\alpha}^{\,\dagger} = {x^\alpha +i\,p^\alpha \over\sqrt2}.
\end{equation}
Besides the energy, $H$, the classical harmonic oscillator has two
additional conserved quanities:  the angular momentum, 
\begin{equation}
M^{\alpha \beta }=x^\alpha \,p^\beta -x^\beta \,p^\alpha 
\end{equation}
and the traceless symmetric tensor
\begin{equation}
S^{\alpha \beta }=x^\alpha \,x^\beta +p^\alpha \,p^\beta -{2\over
m}\,\delta^{\alpha \beta }\, H .
\end{equation}
These conserved tensors correspond, respectively to the antisymmetric
and the symmetric parts of the $SU(m)$ generators,
\begin{equation}
T^\alpha _\beta ={a^\alpha}^{\,\dagger} \, a_\beta
-{1\over m} \delta^{\alpha\beta} {a^\gamma}^{\,\dagger}\, a_\gamma
=\half\,S^{\alpha \beta }-{i\over 2}\,M^{\alpha \beta } .
\end{equation}

The raising operators are tensor operators transforming under $SU(m)$
like the defining $m$ dimensional representation.  The quantum states
are constructed by acting on the ground state with a product of
raising operators.  Since the raising operators commute, a state with
$N$ raising operators transforms as the completely symmetric $SU(m)$
tensor with $N$ indices---the same representation in which arose for 
the large $N$ baryons.  We would expect that for large $N$, the
conserved quantities would approach their classical values. Let us
show that this happens.

The most general classical motion of the $m$ dimensional harmonic 
oscillator is
\begin{equation}
\vec x(t) = \vec v_1\, \cos(t) + \vec v_2\, \sin(t)\quad\quad
\vec p(t) = -\vec v_1\, \sin(t) + \vec v_2\, \cos(t)
\end{equation}
The real vectors $\vec v_1$ and $\vec v_2$ satisfy
\begin{equation}
{\vec v_1}^{\,2} + {\vec v_2}^{\,2}=
{\vec x}^{\,2} + {\vec p}^{\,2} = 2E .
\end{equation}
In the state with principle quantum number $N$, the energy is $N+m/2$, or
approximately $N$ for $N$ large.  Thus, in the large $N$ limit,
\begin{equation}
{\vec v_1}^{\,2}+{\vec v_2}^{\,2} = 2N .
\end{equation}
The traceless tensor $S^{\alpha \beta}$ becomes
\begin{equation}
v_1^\alpha \, v_1^\beta + v_2^\alpha \, v_2^\beta 
- {1\over m}\, \delta^{\alpha\beta}\,
\left({\vec v_1}^{\,2} + {\vec v_2}^{\,2}\right)
\end{equation}
while the angular momentum tensor becomes
\begin{equation}
v_1^\alpha \, v_2^\beta - v_2^\alpha \, v_1^\beta .
\end{equation}
Now introduce the complex vector
\begin{equation}
\sqrt{2N}\, \vec u = \vec v_1 + i\, \vec v_2
\end{equation}
so that
\begin{equation}
u^{\dagger}\, u=1
\end{equation}
In terms of this complex vector $\vec u$, we discover that 
\begin{equation}
T^\alpha_\beta
= {1\over 2}\, S^{\alpha\beta} - {i\over 2}\, M^{\alpha\beta}
= N\, \left( u^\alpha\, u^{\beta\, *} - {1\over m}\,
\delta^{\alpha\beta}  \right)
\end{equation}
which is just (\ref{u}).

\section{The Eigenstates of the Generators\label{eigenstates}}

Our ultimate goal is to describe the leading $N$ behavior of matrix
elements of the $SU(m)$ generators between low spin baryon external
states.  The simplest states, however, with which to describe the action
of the generators are the eigenstates introduced in equation (\ref{u})
since on these eigenstates the group generators act diagonally.  To
connect these two pictures, we must learn how to express the baryons
in terms of these generators.  Since the set of eigenstates is
complete, it should be possible to write any state, such as that for a
baryon or for a harmonic oscillator energy eigenstate, as some
distribution over the space labeling the eigenstates of the
generators.  Conceptually, it is simplest to proceed in two
steps---first to express the harmonic oscillator energy eigenstates in
terms of the $| u \rangle$'s and then to write baryons in terms of
these energy eigenstates.

To begin, we choose a system of coordinates to describe an
eigenstate of the $SU(m)$ generators.  Up to an overall phase, to each
generator corresponds a unique unit vector, $u$; we shall use
this arbitrariness of phase to set the $m^{\rm th}$ component of $u$ to be
purely real.  The remaining components of $u$ can be expressed in terms
of $(m-1)$ positive numbers $w_\alpha$ and corresponding phases
$\theta_\alpha$.  Thus the vector $u(\vec w,\vec\theta)$ becomes
\begin{equation}
\begin{array}{l}
u_\alpha = e^{i\theta_\alpha }\,\sqrt{w_\alpha }\qquad\qquad
[\alpha=1,\ldots, m-1] \\
u_m = \sqrt{1 - \sum_\alpha \, w_\alpha } .
\end{array}
\label{coord}
\end{equation}
Since these coordinates are continuous, the eigenstates require a
continuum normalization,
\begin{equation}
\langle u(\vec w',\vec\theta') | u(\vec w,\vec\theta)\rangle
= \prod_{\alpha=1}^{m-1}\, \delta(w'_\alpha -w_\alpha )\,
\delta(\theta'_\alpha - \theta_\alpha )
\end{equation}
which has the correct $\delta(w'-w)$ dependence to ensure that the
measure is $SU(m)$ invariant.

Having chosen this coordinate system, we next establish the connection
between the harmonic oscillator energy eigenstates and the $u(\vec
w,\vec\theta)$ eigenstates.  The oscillator has correctly normalized
eigenstates of the form
\begin{equation}
| n \rangle \equiv
\left(\prod_{\alpha=1}^m \,{ [{a^\dagger}_\alpha]^{n_\alpha} 
\over\sqrt{n_\alpha !}}\right)
\Ket{0}
\label{hostates}
\end{equation}
where $N=\sum_\alpha \,n_\alpha $ goes to infinity.  Since the
eigenstates are complete, we can generally write $|n\rangle$ as some 
distribution over the space of eigenstates $|u(\vec w,\vec\theta)\rangle$,
specified by a ``spectral function'' $C(\vec n, \vec w, \vec\theta)$:
\begin{equation}
\Ket{n} = \int d\vec w d\vec\theta\, C(\vec n, \vec w, \vec\theta)\,
| u(\vec w,\vec\theta) \rangle + {\cal O}(1/N) .
\label{hostates2}
\end{equation}
Acting upon both sides of this expansion with the group generators the
group generators, $T^\alpha_\beta$, determines the behavior of the
function $C(\vec n, \vec w, \vec\theta)$:
\begin{equation}
\begin{array}{l}
\ds \left( a^{\alpha\,\dagger} \, a_\beta
-{1\over m} \delta^\alpha_\beta a^{\gamma\,\dagger}\, a_\gamma
\right) \Ket{n} = \\
\ds \quad N \int d\vec w d\vec\theta\, \left( u^\alpha \bar u_\beta
-{1\over m} \delta^\alpha_\beta  \right)\,
C(\vec n, \vec w, \vec\theta)\, | u(\vec w,\vec\theta)\rangle
+ {\cal O}(1/N).
\end{array}
\end{equation}
For example, for a diagonal element $T^\alpha_\alpha$ (with a fixed
$\alpha \le m-1$), this equation gives
\begin{equation}
\left( n_\alpha/N - 1/m \right) \Ket{n} =
\int d\vec w d\vec\theta\, \left( w_\alpha - 1/m \right)\,
C(\vec n, \vec w, \vec\theta)\, | u(\vec w,\vec\theta)\rangle
+ {\cal O}(1/N)
\label{hostates3}
\end{equation}
which in turn implies
\begin{equation}
(w_\alpha - n_\alpha /N)\,C(\vec n, \vec w, \vec\theta) \approx 0
\label{hostates4}
\end{equation}
suggesting that the function $C(\vec n, \vec w,\vec\theta)$ peaks
sharply near $w_\alpha = n_\alpha/N$.

Looking at the action of off-diagonal elements of the generators
further constrains $C(\vec n, \vec w,\vec\theta)$ to be of the form
\begin{equation}
C(\vec n, \vec w,\vec \theta)
\approx e^{i\vec\theta\cdot\vec n }\,h(\vec w-\vec n/N) .
\label{hostates5}
\end{equation}
Here, $h(\vec w-\vec n/N)$ describes a function narrowly peaked at the
origin and which satisfies the condition,
\begin{equation}
\int d\vec w\, |h(\vec w-\vec n/N)|^2=\left({1\over2\pi}\right)^{m-1},
\end{equation}
to be consistent with the normalization of the harmonic oscillator states. The
detailed form of the function $h$ will not matter.
Note that the vectors $\vec n$, $\vec w$, and $\vec\theta$ are all 
$(m-1)$ dimensional.

Having determined the form of the function $C(\vec n, \vec w, \vec\theta)$
in equation (\ref{hostates5}), we can compute the matrix element between
two harmonic oscillator energy eigenstates of any finite degree polynomial 
of the generators, $F(T)$ using
\begin{equation}
\begin{array}{l}
\langle n_1,n_2,\ldots | F(T) | n_1+m_1,n_2+m_2,\ldots \rangle \\ \qquad\ds
=\left({1\over2\pi}\right)^{m-1}\,\int\,d\theta\,e^{i\vec\theta\cdot\vec m}
\,F\Bigl({\cal T}(w,\theta)\Bigr)
\end{array}
\label{hostatesresult}
\end{equation}
where the $n_\alpha\rightarrow\infty$ with the $m_\alpha$ and
$w_\alpha=n_\alpha/N$ fixed. Roughly speaking, the magnitudes, $w_\alpha$,
record the gross pattern of the $n_\alpha$'s, but because the $n_\alpha$'s get
divided by $N$ to give $w_\alpha$, this information is not precise enough
enough to distinguish $n_\alpha$ from $n_\alpha+1$ or $n_\alpha+2$, etc.
Those distinctions are produced by the $\theta_\alpha$ dependence, which
records the fine details of the states in a kind of holographic fashion.

\section{Low spin states\label{spin}}

The states that interest us for the application to large $N_c$ baryons
are the low spin states.  For the remainder of this letter, we
restrict to an $SU(4)$ spin-flavor symmetry appropriate for baryons
built only from $u$ and $d$ quarks.  As described in \cite{djm} and 
\cite{luty}, since the baryons always have exactly the same antisymmetric 
behavior in color, we can equivalently examine a baryonic state as
though built from bosonic, colorless quarks which have the have
$SU(m)$ group theoretical properties.

We specify our representation for the quarks as follows:
\begin{equation}
\begin{array}{l}
\mbox{$a^\dagger_1$ creates a $u$ quark with spin up,}\\
\mbox{$a^\dagger_2$ creates a $u$ quark with spin down,}\\
\mbox{$a^\dagger_3$ creates a $d$ quark with spin up,}\\
\mbox{$a^\dagger_4$ creates a $d$ quark with spin down.}
\end{array}
\end{equation}
This assignment has the structure of a tensor product of the two
dimensional flavor space and the two dimensional spin space, and we
can take the spin and isospin generators to be half the usual Pauli
matrices in the appropriate space.

In a low-spin large $N$ baryon, there are a few valence quarks, but most of the
quarks are combined in spin-0 pairs. If there are only two quark flavors, these
spin-0 pairs also have isospin 0 (this is what makes the two-flavor case so much
simpler that three flavors). Thus
up to a normalization, a low spin baryon corresponds to the following
combination of bosonic creation operators:

\begin{equation}
| N,k_\alpha \rangle \propto
z^{\dagger (N-\tilde k)/2}\,
a_1^{\dagger k_1}\, a_2^{\dagger k_2}\,
a_3^{\dagger k_3}\, a_4^{\dagger k_4}\,
| 0 \rangle
\label{lowspin}
\end{equation}
where
the number of valence quarks, $\tilde k=\sum_\alpha \,k_\alpha$, is much smaller
than $N$ and finite as $N\rightarrow\infty$, and
the operator $z^\dagger$
\begin{equation}
z^\dagger \equiv a_1^\dagger\, a_4^\dagger - a_2^\dagger\, a_3^\dagger
\label{z}
\end{equation}
creates a spin and isospin zero pair.
We would like to write $|N,k_\alpha\rangle$ as a sum of harmonic
oscillator eigenstates, since from the previous section, we learned
how to express them in terms of the eigenstates of the $SU(m)$
generators.  We therefore expand ${z^\dagger}^{(N-\tilde k)/2}$, via
the binomial theorem, and use equation (\ref{hostates}) to express
the expansion in oscillator eigenstates:  [$n\equiv N - \tilde k$]
\begin{equation}
\begin{array}{r@{\;}c@{\;}l}
| N,k_\alpha \rangle &\propto&\ds
\sum_{j=0}^{n/2}\,(-1)^j\,{(n/2)!\over j!\,(n/2-j)!}\\
&\cdot&\sqrt{ (n/2-j+k_1)!\, (j+k_2)!\, (j+k_3)!\, (n/2-j+k_4)!}\\
&\cdot& | n/2-j+k_1,\; j+k_2,\; j+k_3,\; n/2-j+k_4\rangle .
\end{array}
\end{equation}
In a typical term, both $n/2-j$ and $j$ are some finite fraction of
$N$---implicitly much larger than any of the $k_\alpha$'s for low
spin baryons---so to leading order this expansion simplifies to 
\begin{equation}
\begin{array}{r@{\;}c@{\;}l}
| N,k_\alpha \rangle&\propto&\ds
(n/2)!\, \sum_{j=0}^{n/2}\, (-1)^j\,
(n/2-j)^{(k_1+k_4)/2}\, j^{(k_2+k_3)/2}\\
&\cdot& | n/2-j+k_1,\; j+k_2,\; j+k_3,\; n/2-j+k_4\rangle .
\end{array}
\label{lowspin2}
\end{equation}

To obtain the matrix elements of the $SU(m)$ generators between low
spin baryons, we can simplify the calculation by directly
evaluating such a matrix element and then determining the
normalization.  The matrix element for a general polynomial $F(T)$ of
the group generators between two baryons specified by the valance
numbers
$k'_\alpha$ and $k_\alpha$ is
\begin{equation}
\begin{array}{l}
\langle N,k'_\alpha | F(T) | N,k_\alpha\rangle \propto \\ 
\quad \ds \sum_{j,\ell=0}^{n/2}\, (-1)^{j-\ell}\,
(n/2-j)^{(k_1+k_4)/2}\, j^{(k_2+k_3)/2} (n/2-\ell)^{(k'_1+k'_4)/2}
\,\ell^{(k'_2+k'_3)/2} \\
\quad \cdot 
\langle {n\over 2}-\ell+k'_1, \ell+k'_2, \ell+k'_3, {n\over 2}-\ell+k'_4
| \\
\quad \cdot F(T) |
{n\over 2}-j+k_1, j+k_2, j+k_3, {n\over 2}-j+k_4\rangle 
\end{array}
\label{lowspinme}
\end{equation}
In terms of the coordinates of the generators (using equations
(\ref{hostates2}) and (\ref{hostates5}) and neglecting the ${\cal O}
(1/N)$ terms, this matrix element becomes
\begin{equation}
\begin{array}{l} \ds
\langle N,k'_\alpha | F(T) | N,k_\alpha\rangle \approx \\
\quad \ds\int\,dw\,d\theta\,
\sum_{j,\ell=0}^{n/2}\,(-1)^{j-\ell}
\,e^{i(j-\ell)(\theta_2+\theta_3-\theta_1)}
\,e^{i\vec\theta\cdot(\vec k-\vec k')}
\,e^{i\theta_1(\tilde k'-\tilde k)/2} \\
\quad \cdot (n/2-j)^{(k_1+k_4)/2}
\,j^{(k_2+k_3)/2}
(n/2-\ell)^{(k'_1+k'_4)/2}
\,\ell^{(k'_2+k'_3)/2} \\
\quad \cdot 
h(w_1-1/2+\ell/N,\,w_2-\ell/N,\,w_3-\ell/N)^{\ds *}\\
\quad \cdot F\Bigl( {\cal T}(w,\theta)\Bigr)\,
h(w_1-1/2+j/N,\,w_2-j/N,\,w_3-j/N) .
\end{array}
\label{lowspinme2}
\end{equation}
In the limit $N\to\infty$, the sum will become an integral.  Since the 
function $h$ sharply peaks at the origin, it imposes that $j/N\approx \ell/N$,
which we will call $x/2\equiv j/N$, as well as that
\begin{equation}
w_1\approx w_4\approx (1-x)/2 \qquad w_2\approx w_3\approx x/2 .
\label{lowspinme5}
\end{equation}
The two sums over $j$ and $\ell$ can then be reorganized into an integral 
over $x$ (from $0$ to $1$) and a sum over $j-\ell$ which becomes
\begin{equation}
\begin{array}{r@{\;}c@{\;}l}
\sum_{j-\ell} (-1)^{j-\ell}e^{i(j-\ell)(\theta_2 + \theta_3 - \theta_1)}
&=& \sum_{j-\ell} e^{i(j-\ell)(\theta_2 + \theta_3 - \theta_1 - \pi)} \\
&\approx& \delta(\theta_2 + \theta_3 - \theta_1 - \pi) .
\end{array}
\label{angles}
\end{equation}
Therefore, in the large $N$ limit, the matrix element becomes
\begin{equation}
\begin{array}{r@{\;}c@{\;}l} \ds
\langle N,k'_\alpha | F(T) | N, k_\alpha \rangle &=&
\ds A_k\,A_{k'}\,\int\,d\theta_1\,d\theta_2\,d\theta_3\,\int_0^1\,dx
\\&\cdot&
e^{i\vec\theta\cdot(\vec k-\vec k')}
\,e^{i\theta_1(\tilde k'-\tilde k)/2}
\,\delta(\theta_2+\theta_3-\theta_1+\pi)\\&\cdot&
(1-x)^{(k_1+k_4+k'_1+k'_4)/2}\cdot
x^{(k_2+k_3+k'_2+k'_3)/2}
\\&\cdot&F\Bigl({\cal T}(w,\theta)\Bigr)
\end{array}
\label{lowspinresult}
\end{equation}
with
\begin{equation}
w_1=w_4=(1-x)/2\,,\quad w_2=w_3=x/2 .
\label{lowspinresult2}
\end{equation}
The remaining normalization factor, $A_k$, is fixed by evaluating the 
integral for $F(T)=1$ to be 
\begin{equation}
A_k={1\over2\pi}\,\sqrt{ (1 + \tilde k)!\over
(k_1 + k_4)!\,(k_2 + k_3)!} .
\label{lowspinresult3}
\end{equation}

An important feature of the integral for the matrix element in
equation (\ref{lowspinresult}) is how it depends on the external
baryons and in particular on the ``valance numbers'' $k_\alpha$ and
$k'_\alpha$.  This dependence occurs in both the $\theta$ terms and in
the $(1-x)^{(k_1+k_4+k'_1+k'_4)/2}\cdot x^{(k_2+k_3+k'_2+k'_3)/2}$
factor which arose from the normalization of the harmonic oscillator
states.  Because of Bose-Einstein statistics, the large $N$ states
depend strongly on the few indices that are not combined into spin and
isospin zero combinations. This is the essential physics of the order
$N$ matrix elements of $T^\alpha_\beta$ between low spin states.

Notice that upon setting $\theta_2+\theta_3-\theta_1 = \pi$, as in
(\ref{angles}), the linear combinations corresponding to the pure spin
or the pure isospin generators vanish identically.  This observation
amounts to the familiar statement that the matrix elements of pure
spin or isospin operators between low spin, large $N$ baryons are only
of order ${\cal O}(1)$, since the integral captures only the leading
behavior in $N$.

\subsection{Some Sample Calculations.}

The integral formula in equation (\ref{lowspinresult}) provides a simple 
means for extracting the leading $N$ features for the low spin baryon matrix
elements.  We will show this simplicity with three examples.  To begin, 
let us calculate the matrix element,
\begin{equation}
\langle p, 1/2 | \sigma_3\tau_3 | p, 1/2 \rangle .
\end{equation}
The spin up proton corresponds to $k_1=1$ with $k_2=k_3=k_4=0$ while the 
operator $\sigma_3\tau_3$ corresponds to the following linear combination 
\begin{equation}
T^{11} + T^{44} - T^{22} - T^{33}
\end{equation}
which becomes $(1-2x)$, with no theta dependence.  
Thus
\begin{equation}
\begin{array}{r@{\;}c@{\;}l}
\langle p, 1/2\, |\, \sigma_3\tau_3\, |\, p, 1/2 \rangle &=& 
\ds {N\over4\pi^2}\, {2!\over 1!\,0!} \int_0^1dx\, (1-x)\,(1-2x)\\
&\cdot& \ds\int d\theta_1d\theta_2d\theta_3\,\,
\delta(\theta_2 + \theta_3 - \theta_1 - \pi) \\
&=&N/3 .
\end{array}
\end{equation}
This result is correct, for the exact answer is 
\begin{equation}
(N+2)/3 .
\label{exact1}
\end{equation}
In fact, this factor is general---the exact large $N$ result for any
matrix
element between spin 1/2 states is simply the $N=1$ result multiplied by
$(N+2)/3$. For $N=3$, this gives the
famous factor of 5/3 for the renormalization of the axial vector current.
Equation (\ref{exact1}) (and
(\ref{exact2general}) and (\ref{exact3general}) below) can be
easily obtained by explicitly constucting the low spin states out of
colorless commuting ``quark'' creation operators, using (\ref{lowspin}) and
(\ref{z}). One can then derive recursion relations relating the matrix
elements for for different values of $N$, and solve them to obtain these
exact results. But if we only need the leading contributions, the integral
formula (\ref{lowspinresult}) captures them all in a much simpler way.

The spin +3/2 $\Delta^{++}$ state corresponds to $k_1=3$, and the matrix
$\sigma_1\tau_1$ corresponds to
\begin{equation}
T^{14} + T^{23} + T^{32} + T^{41} .
\end{equation} 
In the matrix element 
\begin{equation}
\langle \Delta,3/2\, |\, \sigma_1\tau_1\, |\, p, 1/2 \rangle .
\end{equation}
The phase from (\ref{lowspinresult}) is $e^{-i\theta_1}$, so only $T^{14}$
(which has an $e^{i\theta_1}$ dependence) can contribute---the others are
eliminated by the $\theta_j$ integrations.  Then the result is
\begin{equation}
\begin{array}{r@{\;}c@{\;}l}
\langle \Delta, 3/2 | \sigma_1\tau_1 | p, 1/2\rangle &=& 
\ds {N\over4\pi^2}\, \sqrt{2!\over1!\,0!}\, \sqrt{4!\over3!\,0!} 
\int_0^1dx\, (1-x)^2 (1-x)/2 \\
&\cdot&
\ds\int d\theta_1 d\theta_2 d\theta_3\,\, 
\delta(\theta_2 + \theta_3 - \theta_1 - \pi)  \\
&=& N/\sqrt8 .
\end{array}
%\label{}
\end{equation}
The exact result is
\begin{equation}
\sqrt{(N-1)(N+5)/8} .
\label{exact2}
\end{equation}
As with (\ref{exact1}), this result can be easily generalized.  The matrix
element between any spin 1/2 state and any spin 3/2 state is the $N=3$ value
multipled by
\begin{equation}
\sqrt{(N-1)(N+5)/4}\;.
\label{exact2general}
\end{equation}

One more, for good measure---let us calculate
\begin{equation}
\begin{array}{r@{\;}c@{\;}l}
\langle \Delta, 3/2\, |\, \sigma_3\tau_3\, |\, \Delta, 3/2 \rangle &=&
\ds {N\over4\pi^2}\, {4!\over3!\,0!}
\int_0^1 dx\, (1-x)^3 (1-2x) \\
&\cdot& \ds\int d\theta_1 d\theta_2 d\theta_3\,\, 
\delta(\theta_2 + \theta_3 - \theta_1 - \pi) \\
&=& 3N/5 .
\end{array}
\end{equation}
The exact result is
\begin{equation}
3\times (N+2)/5 .
\label{exact3}
\end{equation}
As before, the matrix
element between any two spin 3/2 states is the $N=3$ value
multipled by
\begin{equation}
(N+2)/5 .
\label{exact3general}
\end{equation}

\section{Conclusions\label{conclusions}}

We believe that the integral formula, (\ref{lowspinresult}), in
addition to providing a simple calculational tool, yields some insight
into the nature of the large $N$ enhancement of matrix elements. The
basic physics is Bose-Einstein statistics. The low spin large $N$
states contain a large number of spin and isospin zero pairs in
addition to the ``valence'' quarks.  In these pairs, because of
Bose-Einstein statistics, the creation operators that duplicate those
of the valence quarks dominate over those that do not appear in the
valence sector. It is this asymmetry that produces the large $N$
matrix element enhancement.

\section*{Acknowledgements}

We are grateful to S.L. Glashow, N. Nekrasov and M. Voloshin for useful
conversations.

\def\PRL#1#2#3{{Phys.\ Rev.\ Lett.}\ {\bf #1}{ (#2) }{#3}}
\def\PRD#1#2#3{{Phys.\ Rev.}\ {\bf D#1}{ (#2) }{#3}}
\def\PLB#1#2#3{{Phys.\ Lett.}\ {\bf #1B}{ (#2) }{#3}}
\def\NPB#1#2#3{{Nucl.\ Phys.}\ {\bf B#1}{ (#2) }{#3}}

\end{document}